\documentstyle[12pt,epsfig]{article}
\let\section=\subsection     \let\subsection=\subsubsection                

\begin{document}

\begin{center}
{\bf INSTITUT~F\"{U}R~KERNPHYSIK,~UNIVERSIT\"{A}T~FRANKFURT}\\
D - 60486 Frankfurt, August--Euler--Strasse 6, Germany
\end{center}

\hfill IKF--HENPG/1--97

\vspace{1cm}

\begin{center}
{\Large \bf PION AND STRANGENESS PRODUCTION }
\end{center}
\begin{center}
{\Large \bf  AS }
\end{center}
\begin{center}
{\Large \bf SIGNALS OF QCD PHASE TRANSTION }

\vspace{1cm}

Marek Ga\'zdzicki\footnote{E--mail: marek@ikf.uni--frankfurt.de}\\
Institut f\"ur Kernphysik, Universit\"at Frankfurt \\
August--Euler--Strasse 6, D--60486 Frankfurt, Germany\\[0.8cm]

\vspace{2cm}

\begin{minipage}{14cm}
\baselineskip=12pt
\parindent=0.5cm
{\small 
It is shown that data on pion and strangeness production 
in central nucleus--nucleus collisions are consistent with the
hypothesis of a Quark Gluon Plasma formation
between 15 A$\cdot$GeV/c
(BNL AGS) and 160 A$\cdot$GeV/c (CERN SPS) collision energies.
The experimental results interpreted in 
the framework of a  statistical approach
indicate that
the effective number of degrees of freedom increases by a 
factor of about 3 in the course of the phase transition
and that the  plasma created at CERN SPS energy 
may have a temperature
of about 280 MeV (energy density $\approx$ 10 GeV/fm$^{3}$).
Experimental studies of central Pb+Pb collisions in the 
energy range 20--160 A$\cdot$GeV/c  are urgently 
needed in order to localize the threshold energy,
and study the properties of the QCD phase transition.
}

\end{minipage}

\vspace{1.5cm}
{\it Talk given at the Internatinal Conference on QCD Phase Transitions,
Hirschegg, Austria, January 13--18, 1997}

\end{center}

\vfill
\today
\newpage

\begin{center}
   {\large \bf PION AND STRANGENESS PRODUCTION AS }\\[2mm]
   {\large \bf SIGNALS OF QCD PHASE TRANSTION }\\[5mm]
   MAREK GA\'ZDZICKI \\[5mm]
   {\small \it  Institut f\"ur Kernphysik, Universit\"at Frankfurt \\
   August--Euler Str. 6, D--60486 Frankfurt/M, Germany \\[8mm] }
\end{center}

\begin{abstract}\noindent
It is shown that data on pion and strangeness production 
in central nucleus--nucleus collisions are consistent with the
hypothesis of a Quark Gluon Plasma formation
between 15 A$\cdot$GeV/c
(BNL AGS) and 160 A$\cdot$GeV/c (CERN SPS) collision energies.
The experimental results interpreted in 
the framework of a  statistical approach
indicate that
the effective number of degrees of freedom increases by a 
factor of about 3 in the course of the phase transition
and that the  plasma created at CERN SPS energy 
may have a temperature
of about 280 MeV (energy density $\approx$ 10 GeV/fm$^{3}$).
Experimental studies of central Pb+Pb collisions in the 
energy range 20--160 A$\cdot$GeV/c  are urgently 
needed in order to localize the threshold energy,
and study the properties of the QCD phase transition.

\end{abstract}

\section{Introduction}

In this paper we review the recent status of our search for a QCD transition
to the Quark Gluon Plasma \cite{qgp} by a systematic analysis of
entropy and strangeness production in nuclear collisions.
There are important reasons to select entropy \cite{Va:82} 
and strangeness \cite{Ko:86}
as basic observables.
Both are defined in any form of strongly interacting matter.
Their equilibrium values are directly sensitive to matter
properties: effective number of degrees of freedom and
their effective masses.
Entropy and strangeness production are also believed to freeze--out at the
early stage of  evolution of a system  created in nuclear 
collisions.

Our strategy of data\footnote{ Data obtained by about 100
different experiments are used. The references to the original
experimental works can be found in the compilation papers \cite{Ga}.} 
analysis \cite{Ga} reviewed here
can be summarized as follows:\\
1. We study the dependence of the properly normalized entropy
(mainly determined by pion multiplicity) and
strangeness (mainly determined by kaon and hyperon yields)
production on the volume of the colliding nuclear matter at a
fixed collision energy.
We demonstrate that a fast saturation occurs.    
The simplest qualitative interpretation is, that equilibration of
entropy and strangeness takes place (Section 2). \\
2. We study the dependence of the saturation levels on the collision
energy.
We demonstrate that the saturation levels for both entropy and strangeness
show an unusual change between AGS ($\approx$15 A$\cdot$GeV/c) and SPS
($\approx$160 A$\cdot$GeV/c) energies.
The simplest qualitative interpretation is,
that the threshold energy for QGP creation is located in the above energy
range (Section 3). \\
3. Finally, we formulate a simple statistical model for entropy and
strangeness production in nuclear collisions and demonstrate that
the experimental results at SPS energy can be quantitatively described
assuming creation of QGP (Section 4).

\section{Volume Dependence}

\begin{figure}[t]
\epsfig{file=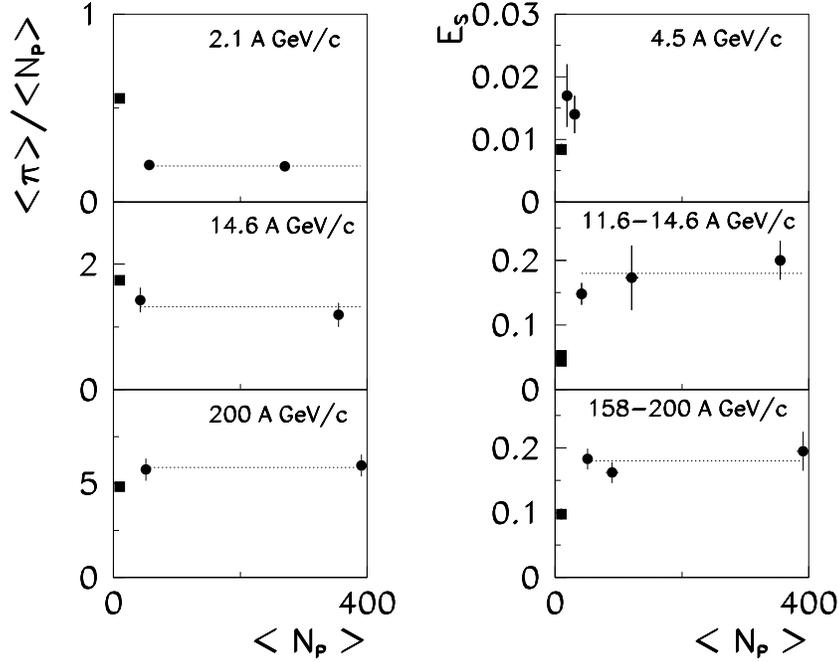,height=10cm}
\caption{ \protect\begin{small}
(left)
 The dependence of the ratio $\langle \pi \rangle/\langle N_P \rangle$
on $\langle N_P \rangle$ at three different collision energies.
(right)
The dependence of the $E_S$ ratio 
on $\langle N_P \rangle$ at three different collision energies.
The data for central A+A collisions are indicated by circles and
the data for N+N interactions by squares.
\protect\end{small} }
\label{fig1}
\end{figure}

Experimental data on pion\footnote{Here we use pion multiplicity instead
of entropy in order to start analysis from `raw' experimental data.
An improved estimate of entropy is given in the next section.}
 and strangeness production in central
nucleus--nucleus (A+A) and all inelastic nucleon--nucleon (N+N)
collisions are shown in Fig. 1  as a function of the number
of participant nucleons, $\langle N_P \rangle$, for various
collision energies.
In order to eliminate a trivial volume dependence, the normalized
multiplicities are studied:
$\langle \pi \rangle/\langle N_P \rangle$
and
$E_S \equiv
( \langle \Lambda \rangle + \langle K 
+ \overline{K} \rangle )/\langle \pi \rangle$,
where $\langle \pi \rangle$, 
$\langle \Lambda \rangle$, and
$\langle K+\overline{K} \rangle$ are
mean multiplicities of pions, $\Lambda/\Sigma^0$ hyperons, and
kaons and antikaons, respectively.
For all energies a similar behaviour is observed:
a rapid change between results for N+N interactions and  intermediate mass
nuclei ($\langle N_P \rangle \approx$ 50) is followed by a well defined
region in which the 
normalized pion and strangeness production is almost constant.
We interpret the observed saturation as a result of an equilibration of entropy
and strangeness yields.

\section{Energy Dependence}

\begin{figure}[t]
\epsfig{file=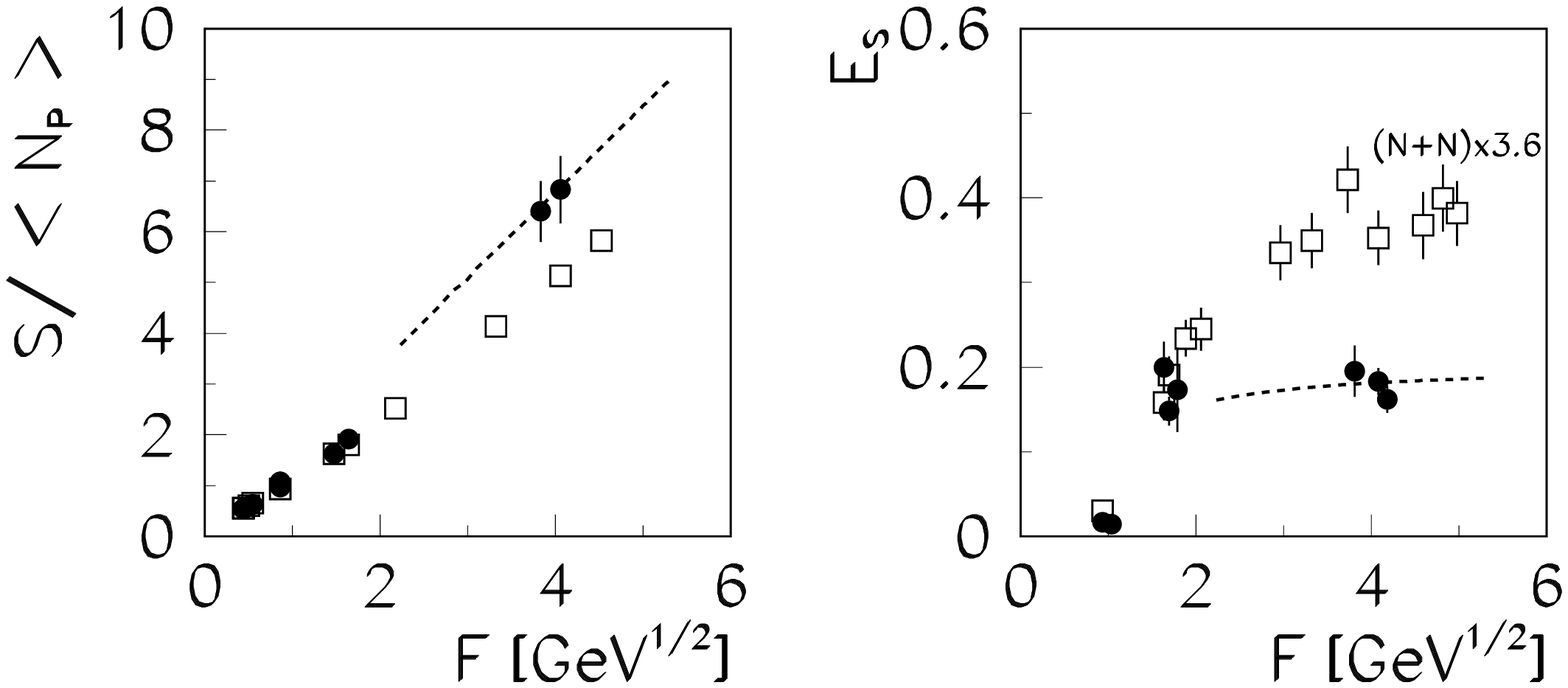,height=6cm}
\caption{\protect\begin{small}
The dependence of  $S/\langle N_P \rangle$ (left) and $E_S$ (right)
on the collision energy measured by the Fermi variable $F$. 
The data for central A+A collisions are indicated by circles and the
data for N+N interactions by squares (the $E_S$ values for N+N interactions are
scaled by a factor of 3.6).
The dashed lines show results obtained within the generalized Landau model
assuming QGP creation.
\protect\end{small}}
\label{fig2}
\end{figure}

The collision energy dependence of the normalized entropy and strangeness
production is shown in Fig. 2.
The energy dependence is studied using the Fermi energy
variable \cite{Fe:50,La:53}
$F \equiv (\sqrt{s}_{NN} - 2 m_N)^{3/4}/\sqrt{s}_{NN}^{1/4}$,
where $\sqrt{s}_{NN}$ is the c.m. energy for a nucleon--nucleon pair and
$m_N$ is the nucleon mass.
There are several advantages in using $F$ as an energy variable.
The measured mean pion multiplicity in N+N interactions 
is approximately proportional
to $F$ \cite{Ga,Go:89}.
In the Landau model \cite{La:53} both the entropy and
the initial temperature of the matter (for
$\sqrt{s}_{NN} \gg 2 m_N$) are also proportional to $F$.

The `entropy' presented in Fig. 2 is calculated as:
$S \equiv \langle \pi \rangle + \kappa \langle K+\overline{K} \rangle +
                               \alpha \langle N_P \rangle$,
where two last components take into account kaon production and pion
absorption \cite{Ga}.
Thus $S$ can be treated as the 
inelastic entropy measured in the pion entropy
units.

The normalized `entropy' for central A+A
collisions  at low energies (AGS and below) follows the dependence 
established by N+N data.
The normalized `entropy' for A+A collisions at SPS energy 
(the data of NA35 and NA49 Collab.) is about 30\%
higher than the corresponding value for N+N interactions. 

The  energy dependence of the $E_S$ ratio is also shown in Fig. 2.
The results for N+N interactions are scaled to fit A+A data at AGS. 
A monotonic increase of $E_S$ between Dubna energy 
(p$_{LAB}$ = 4.5 A$\cdot$GeV/c) and SPS energy 
is observed.
In the range from AGS to SPS  the $E_S$ ratio
for N+N interactions is enhanced by a factor of about 2.
A qualitatively different  dependence  is
seen for central A+A collisions.
The rapid increase of  $E_S$ between Dubna and AGS energies
is followed by a weak change of  $E_S$ between AGS and  SPS
collision energies.

Let us now try to understand the observed energy dependence on a
qualitative level.
In the generalized Landau model \cite{Ga} the inelastic entropy
is proportional to
$g^{1/4} \langle N_P \rangle F$,
where $g$ is the effective number of degrees of freedom.
Thus the observed deviation of the data for A+A collisions 
from the Landau scaling, $S/\langle N_P \rangle \sim F$,
can be interpreted as due to an increase of the effective number of 
degrees of freedom when crossing the transition collision energy. 
The magnitude of this increase can be estimated,
within the model, as the fourth power of the ratio of 
slopes of straight lines describing  
low and high energy A+A data:
1.33$^4$ $\approx$ 3 \cite{Ga}.

The second dominant effect of the transition 
to QGP is the reduction of the effective masses of degrees of
freedom.
Basic thermodynamics tells us that for massless particles 
the ratio {\it particle number/entropy}  is independent of the temperature.
For massive particles the ratio increases with $T$ at low
temperature and approaches the  saturation level (equal to the 
corresponding ratio for massless particles) at high temperatures,
$T \gg m$.
This property can be used to study the magnitude of the 
effective mass
of strangeness carriers in strongly interacting matter.
The $E_S$ ratio is approximately proportional to the ratio 
{ \it number of strangeness carriers/entropy (strangeness/entropy)} 
and therefore
its temperature (collision energy, $F$) dependence should be sensitive
to the effective mass of strangeness carriers. 
Reducing  the mass of
strangeness carriers should cause a weaker dependence of
the $E_S$ ratio on the collision energy.   
The rapid increase of the $E_S$ ratio in the energy 
range of $ F < $ 2 GeV$^{1/2}$ 
can be interpreted as due to the large effective mass of strangeness
carriers
(kaons or constituent strange quarks, $ m_S \approx $ 500 MeV/c$^2$ )
in comparison to the temperature of matter, $T < T_C \approx $ 150 MeV.
At temperatures above $T_C$, the matter is in the form of QGP
and the mass of strangeness carriers is equal to the mass of
current strange quarks, $ m_S \approx $ 150 MeV/c$^2$, consequently
$ m_S \leq T $.
Thus a much weaker dependence of the $E_S$ ratio on $F$ is expected 
in the high energy region where the creation of QGP takes place.
The equilibrium value
of the {\it strangeness/entropy} ratio
is higher in hadronic matter (HM) than in QGP at very high temperatures
\cite{Ka:86}.
This is due to the fact that it is proportional
to the ratio of the effective number of strangeness degrees of freedom
to the number of all degrees of freedom.
This ratio is lower in QGP than in HM.
At low temperature, however, the {\it strangeness/entropy}
ratio is lower in HM than in QGP.
This is caused, as previously discussed, by the high mass of strangeness
carriers in HM. 
Thus, in general, a transition to QGP may lead to an increase or a decrease
of the {\it strangeness/entropy} ratio depending at which temperatures
of QGP and HM the  comparison is made.

The presented data suggest that the transition is associated with a
decrease of the {\it strangeness/entropy} ratio.

\section{Model of QGP in A+A}

Encouraged by the qualitative agreement of the data with the  hypothesis
of the equilibrium QGP creation in the early stage of A+A collision at SPS, 
we attempt to make a quantative comparison using  
the generalized Landau model.
We assume that inelastic energy is deposited in a volume equal to the
Lorentz contracted  volume of a nucleus (we consider collisions of two
identical nuclei only). In this volume an equilibrated QGP is formed.
For  simplicity, we assume that the created QGP is baryon 
free\footnote{Introduction of baryon rich QGP significantly complicates
a model. We checked, however, that it seems to have only small
influence on the final results.}.
Furthermore,
we assume that inelastic entropy and strangeness are not changed
during  the system evolution.
The inelastic energy at SPS was measured to be (67$\pm$7)\% of
the available energy \cite{Ba:94}, the same for S+S and Pb+Pb collisions.
It is also weakly dependent on the collision energy  between 
AGS and SPS \cite{St:96}.
The effective radius of the nucleus was taken to be 
$r_0 A^{1/3}$ with $r_0$ = 1.30$\pm$0.03 fm given by the
fit to the inelastic cross section data \cite{cs}.
The QGP is assumed to consist of  gluons, $u, d$, and $s$
quarks, and the corresponding antiquarks.
The strange quark mass was taken to be 175$\pm$25 MeV/c$^2$ \cite{Le:96}. 
This mass is determined at an energy scale of 1 GeV. 
The conversion factor between the calculated entropy and the `entropy'
evaluated from the experimental data is taken to be 4
(entropy per pion at T $\approx$ 150 MeV).
The conversion factor between total strangeness and strangeness measured
by the sum of $\Lambda$ and $K+\overline{K}$ yields is taken to be
1.36 according to the N+N data and a procedure developed in \cite{Bi:92}.
It should be stressed that the model formulated in the above way
has   no free parameters.

The resulting production of entropy and strangeness in the energy range
30--500 A$\cdot$GeV is represented by dashed lines in Fig. 2.
The agreement with the data is surprisingly good.
The analysis suggests that plasma created at the SPS  has an energy 
density of about 10 GeV/fm$^3$ and a temperature of about 280 MeV.

\section{Summary and Conclusions}
 
The experimenal data on pion and strangeness production indicate:\\
-- saturation of pion and strangeness production with the
number of participant nucleons,\\
-- change in the collision energy dependence taking place between
15 A$\cdot$GeV/c and 160 A$\cdot$GeV/c.\\
Within a  statistical approach
the observed behaviour can be qualitatively understood as due to:\\
-- equilibration of entropy  and strangeness  in
collisions of heavy nuclei,\\
-- transition to QGP occuring between
AGS and  SPS energies 
associated with the increase of the effective number of degrees of freedom
by a factor of about~3.\\
\noindent
These observations hold already for the central S+S collisions,
they are not unique to central Pb+Pb collisions.

The results at SPS energy are in surprisingly good agreement 
with the calculations done within the 
generalized Landau model.
The analysis suggests that plasma created at SPS has a temperature of
about 280 MeV (energy density of about 10 GeV/fm$^3$).

Experimental studies of central Pb+Pb collisions in the energy range
20--160 A$\cdot$GeV are {\bf urgently} needed in order to localize 
the threshold energy more precisely and study the properties
of QCD phase transition.

{\it Acknowledgements.} I would like to thank 
Stanis{\l}aw Mr\'owczy\'nski and  J\"org G\"unther for comments
to the manuscript.


\begin{thebibliography}{99}
\itemsep=0cm

\bibitem{qgp} J. C. Collins and M. J. Perry, Phys. Rev. Lett.
{\bf 34} (1975) 151,
E. V. Shuryak, Phys. Rep. {\bf C61} (1980)
71 and {\bf C115} (1984) 151.
\bibitem{Va:82} L. Van Hove, Phys. Lett. {\bf B118} (1982) 138.
\bibitem{Ko:86} P. Koch, B. M\"uller and J. Rafelski, Phys. Rep. 
{\bf 142} (1986) 321.
\bibitem{Ga} M. Ga\'zdzicki and D. R\"ohrich, Z. Phys. 
{\bf C65} (1995) 215,
M. Ga\'zdzicki, Z. Phys. {\bf C66} (1995) 659,
M. Ga\'zdzicki and D. R\"ohrich,
Z. Phys. {\bf C71} (1996) 55,
M. Ga\'zdzicki, Frankfurt University Preprint IKF--HENPG/2--96,
hep--ph/9606473,
M. Ga\'zdzicki, M. I. Gorenstein and St. Mr\'owczy\'nski,
Frankfurt University Preprint IKF--HENPG/6--96, nucl-th/9701013.
\bibitem{Fe:50} E. Fermi, Prog. Theor. Phys. {\bf 5} (1950) 570.
\bibitem{La:53} L. D. Landau, Izv. Akad. Nauk SSSR, Ser. Fiz.
{\bf 17} (1953) 51,\\
S. Z. Belenkij and L. D. Landau, Uspekhi Fiz. Nauk {\bf 56}
(1955) 309.
\bibitem{Go:89} A. I. Golokhvastov, Dubna Report, JINR E2--89--364
(1989).
\bibitem{Ka:86} J. Kapusta and A. Mekjan, Phys. Rev. {\bf D33}
(1986) 1304.
\bibitem{Ba:94} J. B\"achler et al., Phys. Rev. Lett. 72 (1994) 1419.
\bibitem{St:96} H. Str\"obele, in Proceedings of Quark Matter 96,
note that the values of inelasticity for S+S and Pb+Pb collisions 
at SPS are plotted
incorrectly~\cite{Ba:94}.
\bibitem{cs} M. Kh. Anikina et al., Dubna Report 1--82--553 (1982),
E. O. Abdurakhimov et al., Z. Phys. {\bf C5} (1980) 1,
T. F. Hoang, B. Cork and H. J. Crawford, Z. Phys. {\bf C29} (1985) 611,
A. Bamberger et al., Phys. Lett. {\bf 205}~(1988)~583.
\bibitem{Le:96} H. Leutwyler, Phys. Lett. {\bf B378} (1996) 313.
\bibitem{Bi:92} H. Bia{\l}kowska et al.,
Z. Phys. {\bf C55} (1992) 491.


\end{thebibliography}
\end{document}